\documentclass[
 aip,
 apl,
 floatfix,
 jmp,
 amsmath,amssymb,
 reprint,%
]{revtex4-1}

\usepackage{stmaryrd}% special character
\usepackage{graphicx}% Include figure files
\usepackage{dcolumn}% Align table columns on decimal point
\usepackage{bm}% bold math
\usepackage{textcomp}%for degree
\usepackage{subfigure}
\usepackage{psfrag}
\usepackage{multirow,amssymb,amsbsy,amsmath}
\usepackage{ctable}
\usepackage{colortbl}
\usepackage{CJK}
\usepackage{amsmath}

\begin{document}

\title{Elongation of energy exchange between femtosecond laser pulses via plasma formation in air }

\author{Zuoye Liu}
\author{Yu Cao}
\author{Yanchao Shi}
\author{Mingze Sun}
\author{Pengji Ding}
\author{Zeqin Guo}
\author{Bitao Hu}
\email[Author to whom correspondence should be addressed. Electronic mail: ]{hubt@lzu.edu.cn}
\affiliation{School of Nuclear Science and Technology, Lanzhou University, 730000, China}

\date{\today}

\begin{abstract}
We experimentally demonstrate energy exchange between a delay-tuned femtosecond beam and two delay-fixed ones as they spatiotemporally overlapped and experienced filamentation in air. The energy exchange process in the relative time delay is dramatically elongated up to 40 $ps$ in the presence of plasma grating, indicating that filamentary beams coupling may be an effective method for filament control. 
\end{abstract}

\pacs{52.38.Hb, 42.65.Jx, 42.60.Hw} 

%\keywords{Energy exchange, plasma grating, femtosecond pulse filamentation, rotational Raman effect, plasma formation}   
% Use showkeys class option if keyword

\maketitle

% \section{Introduction}
Femtosecond laser pulse filamentation in air is an interesting phenomenon \cite{A.Braun, OL.36.463, PRL.98, Science.301.61, AIP.Advance.2, J.Quantum.Electron, OE.20}, and has intensive applications including few-cycle pulse generation, THz radiation, remote-sensing of atmospheric pollution, lighting and discharge triggering. Recently, much attention has been given to the interaction among few noncollinear crossing filaments. Because of laser field interference, an one- or two- dimensional plasma grating can be formed in the intersection region when the noncollinear filamentary pulses overlap in time and space. Energy exchange between two filamentary pulses has been demonstrated in several experiments including filamentation in air and liquid methanol \cite{PRL.102, PRL.105, OL.37, APL.101, PRL.103}, and the formation of plasma grating plays an important role in this coupling process. For instance, the traveling plasma grating formed at the intersection of two filamentary pulses with slightly different central frequency is responsible for an efficient energy exchange between the filaments \cite{PRL.105}. It has been found that the direction of energy transfer depends on the relative time delay between filamentary pulses, the initial chirp, the laser intensities, the location, the intersecting angle and the relative polarization. However, the underlying physical mechanisms of energy exchange in literatures are different and puzzling, including the impulsive Raman nonlinear response of the molecules \cite{PRL.102}, the plasma-mediated forward stimulated Raman scattering \cite{PRL.103}, the traveling plasma grating \cite{PRL.105}, the configuration akin produced by the crossed filaments to coupled waveguides \cite{OL.37} and the classical two beam coupling model \cite{APL.101}. Bernstein \textit{et al} obtained a conclusion that no significant affect of the filament formation on the energy exchange \cite{PRL.102}, but later Y. Liu \textit{et al} found that the direction of energy exchange reverses due to filament formation when the laser power $P = 3.5 P_{cr}$ \cite{PRL.105}. Besides, all these works were done by using two filamentatary pulses interacting configuration. 

In this letter, we experimentally studied the filamentary pulse coupling between a delay-tuned beam and two delay-fixed beams, keeping in mind underlying physical mechanisms of energy exchange. Our results are consistent with those of previous works demonstrating energy exchange between two filaments in air and liquid methanol, when the relative time delay between the delay-tuned beam and the two delay-fixed ones is varied among approximately 1 ps. Unexpectedly, we found a previously undocumented energy exchange with approximate 40 ps time-scale in the relative time delay.

%\section{Experimental setup}
\begin{figure*}[tb!] 
\centering
\subfigure{
\includegraphics[width=2.2in]{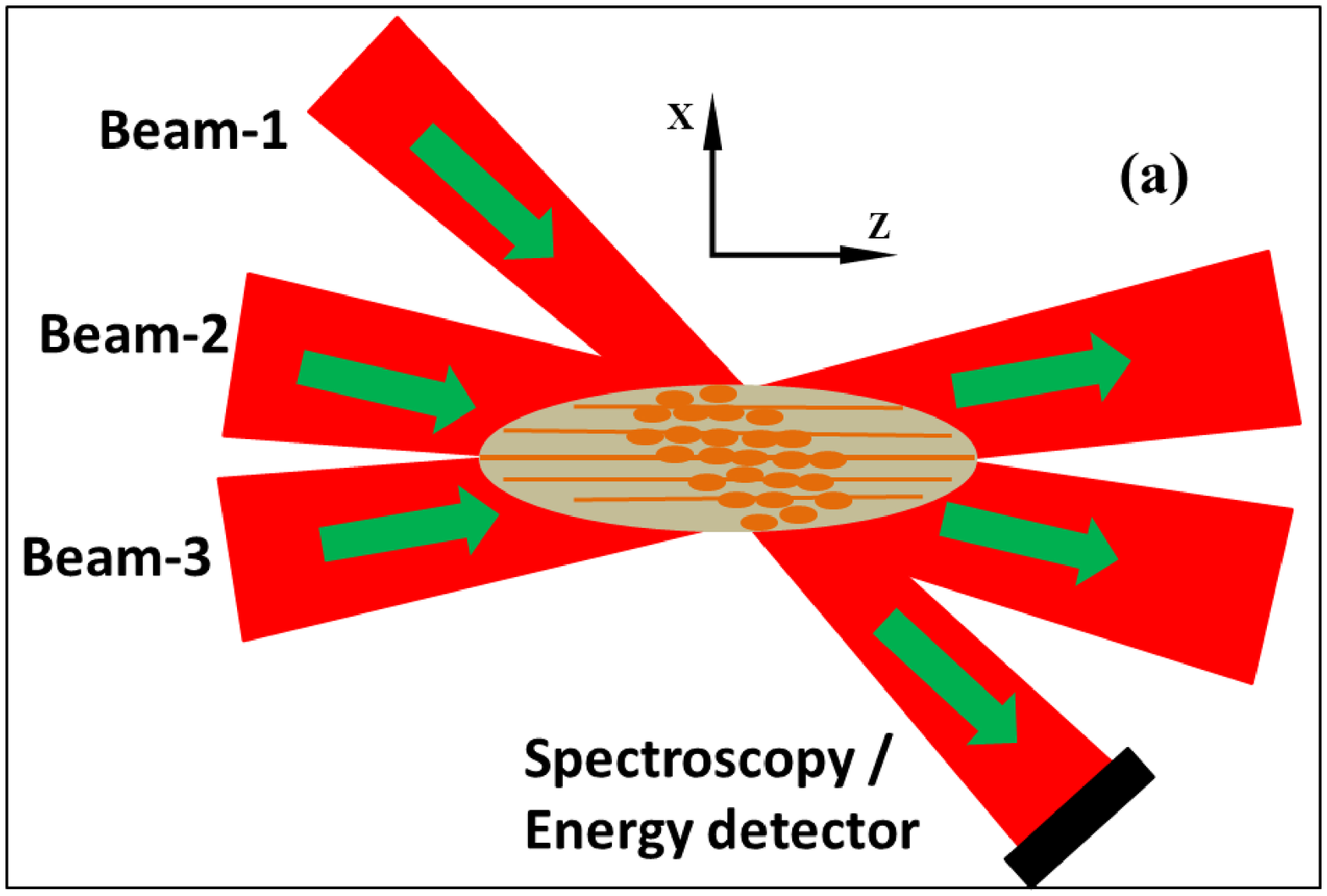}}
\subfigure{
\includegraphics[width=2.2in]{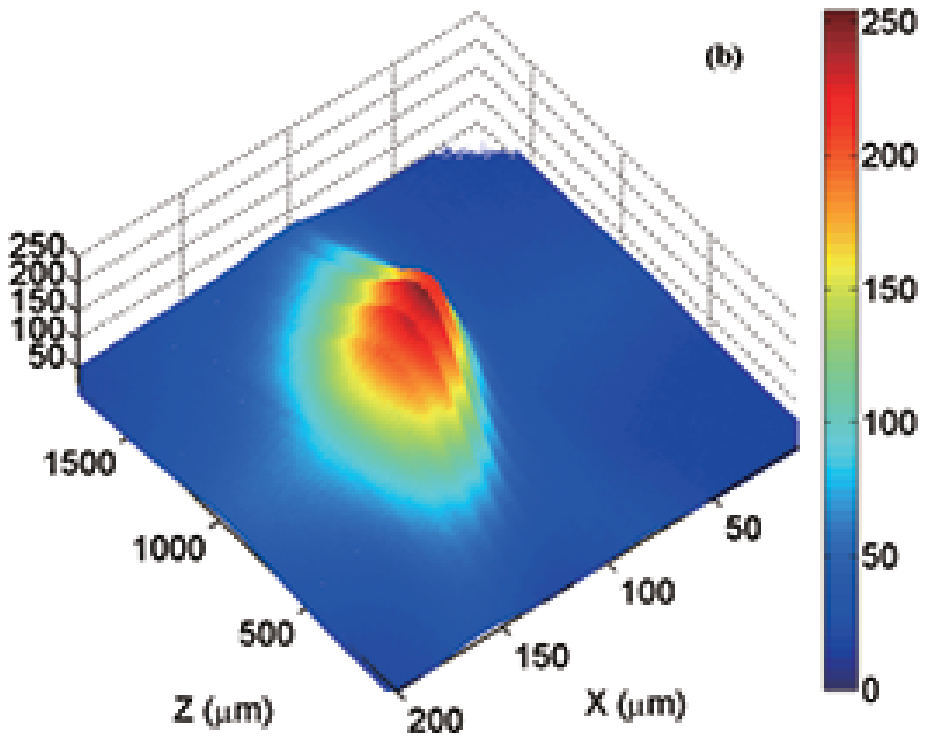}}
\subfigure{
\includegraphics[width=2.2in]{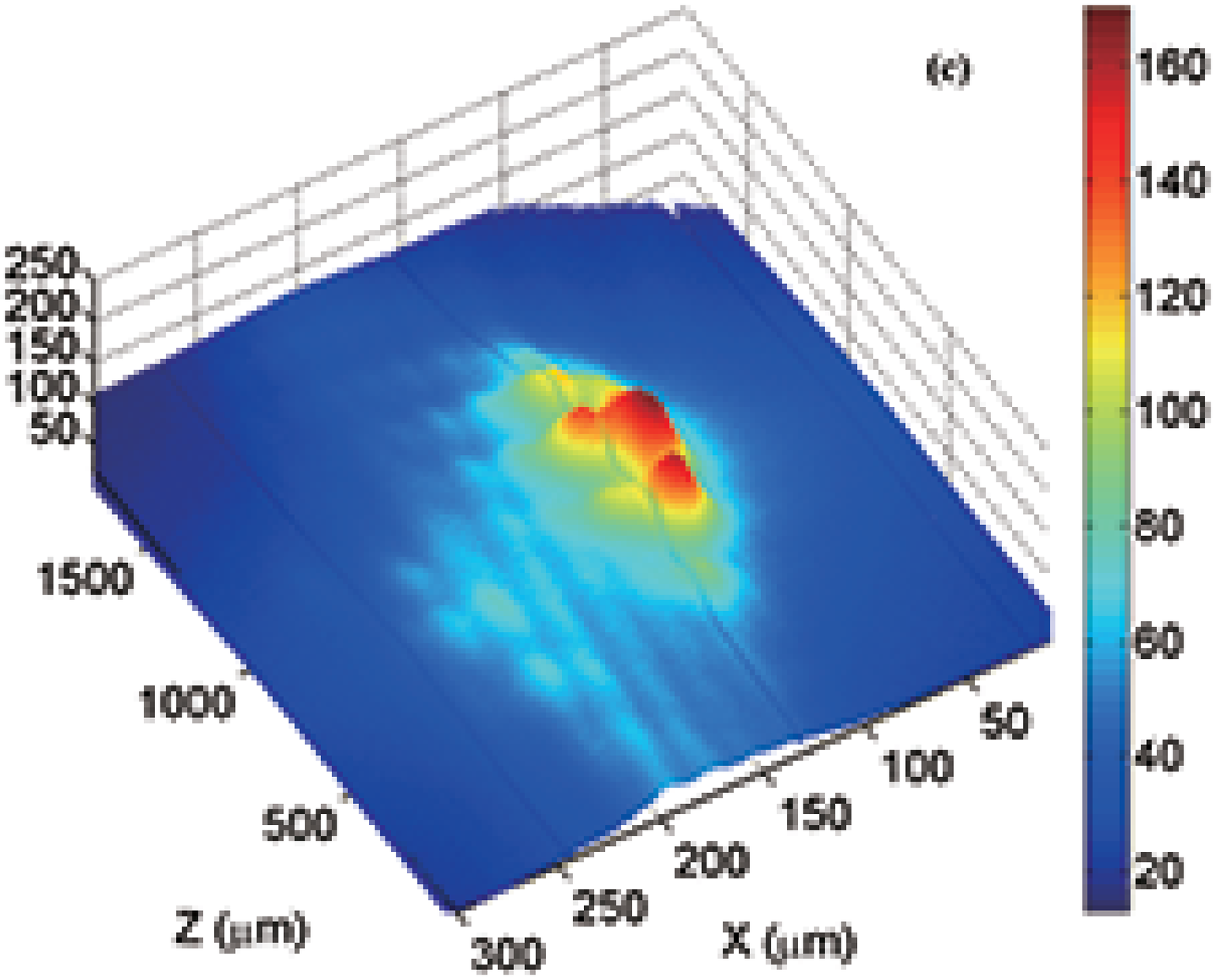}}
\caption{(Color online) (a) Schematic illustration of three filaments interaction. (b) CCD image of one-dimensional plasma grating generated by two delay-fixed filaments when the spatiotemporal overlap is fulfilled. (c) CCD image of the volume plasma grating generated by the delay-tuned filament and two delay-fixed filaments.}
 \label{FIG-1}
\end{figure*}

The interaction of three filaments is schematically shown in Fig .1. The laser system we used is a Ti:sapphire amplifier system which is capable of producing 33 fs pulses centered at 810 nm, at a repetition rate of 1 kHz. The laser pulses are chirp-free, which can be confirmed by the FROG measurements. The laser beam was split into three arms by two successive beam-splitters. One arm with pulse energy of 0.6 mJ, which was defined as Beam-1, and the other two arms defined as Beam-2 and Beam-3 with pulse energies up to 0.8 mJ and 0.7 mJ respectively, were focused by lens. The minimal translation setup corresponds to a temporal resolution of 16.7 fs. The energy of each beam was measured by an energy monitor. The similar details of experimental setup can refer to our previous work \cite{PRA.87}.

% \section{Results and Discussion}
In the experiment a plasma grating was formed when the two delay-fixed pulses without central frequency difference were spatiotemporal overlapped. Because of the plasma density modulation in the plasma grating, the nonlinear refractive index originated from the formation of plasma $\Delta n_{plasma} \sim -\rho (\overrightarrow{r}, t) / \rho _{cr}$ changes periodically, where $\rho _{cr} = 1.74 \times 10^{21} cm^{-3}$ is the critical plasma density for $810 \ nm$ laser pulse in air \cite{Phys.Rep}. The period of this one-dimensional plasma grating can be expressed by $\Lambda = \lambda / [2 sin(\theta / 2)]$, where $\lambda $ and $\theta = 2.9^{\circ }$ are the wavelength of the incident delay-fixed pulses and their crossing angle respectively. For $2.9^{\circ }$ crossing geometry of two delay-fixed beams, the one-dimensional grating period could be estimated as $\Lambda_{1D} \approx 15.8 \ \mu $m which is close to our experiment results in Fig. 1(b), and its refractive index planes would be approximately oriented along the bisector of the angle formed by the two delay-fixed beams. With the presence of the delay-tuned beam with a crossing angle $3.6^{\circ }$ with respect to Beam-2, a volume plasma grating was formed when the three beams tempoprally overlapped, as shown in Fig. 1(c). As a result, its orientation was slightly deviated from that of one-dimensional plasma grating and tended to the propagation direction of the delay-tuned beam. Since the involved laser pulses possess the equal central frequence, their interference field will lead to the formation of the relatively stationary plasma grating rather than a traveling one.
\begin{figure}[ht]
  \centerline{\includegraphics [clip, width=3.5in, angle=0]{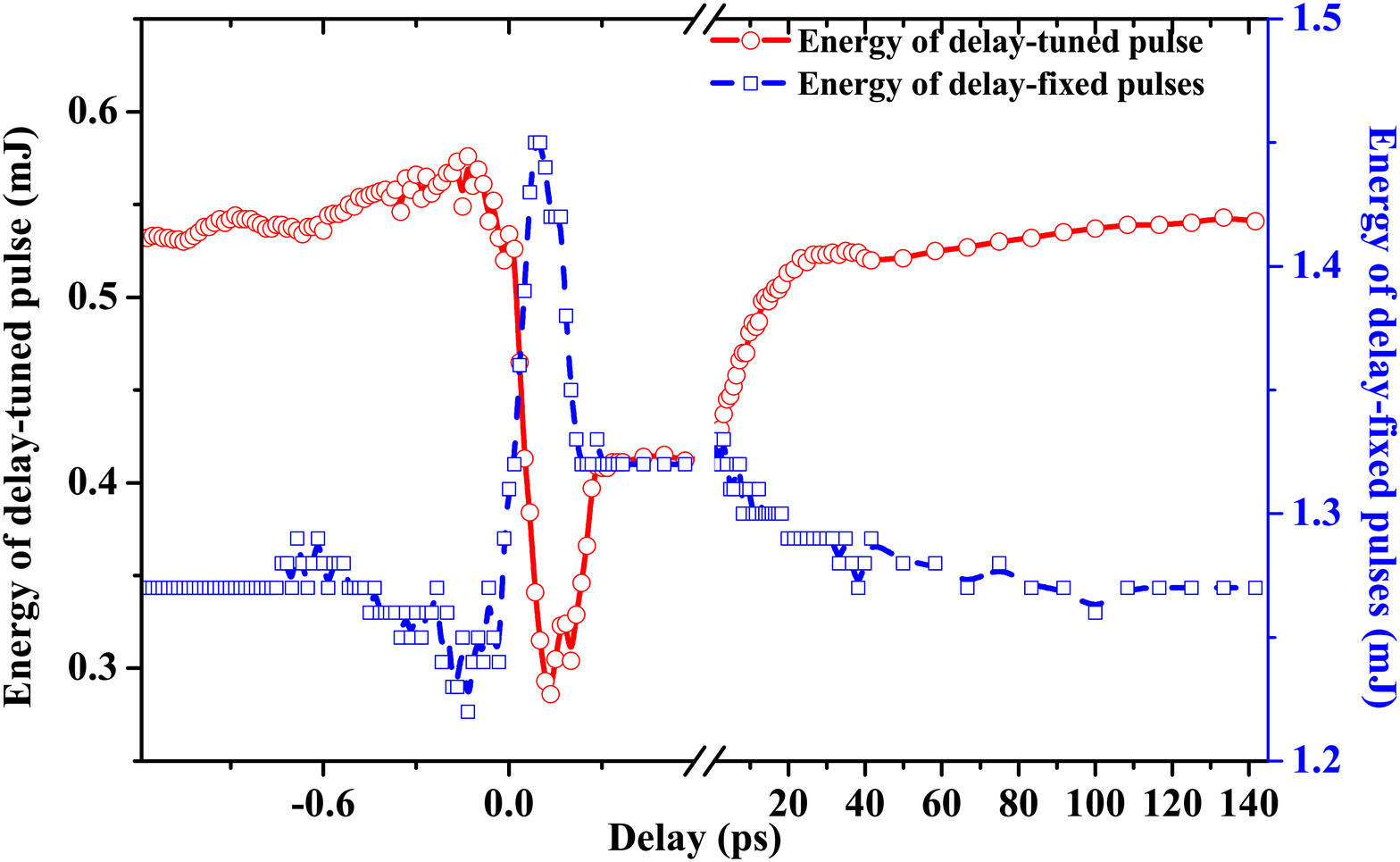} }
  \caption{(Color online) The energies of delay-tuned pulse (circle curve) and two delay-fixed ones (square curve) as a function of the relative time delay. The initial energy of the delay-tuned pulse is 0.53 mJ, while the initial energy of the two delay-fixed pulses is 1.27 mJ.}
  \label{FIG-2}
\end{figure}

Figure 2 shows the energies of the delay-tuned pulse and two delay-fixed ones as a function of their relative time delay. The positive delay corresponds to the translation-stage delay-tuned beam delayed with respect to the two delay-fixed beams. There are two distinct regimes, which is consistent with reported observations \cite{PRL.105, PRL.102, APL.98, APL.101}. The two delay-fixed beams transfer energy to the delay-tuned one during the negative delay, while the delay-tuned pulse transfers energy to the two delay-fixed ones as the relative time delay is positive. It is similar to the observation by Y. Liu \textit{et al} and clearly different from that by Bernstein \textit{et al} in which the trailing pulse obtains energy from the leading pulse. Note that the energy reduction of the delay-tuned pulse is larger than that transfered to the two delay-fixed pulses during the pulses overlap time because of the nonlinear multiphoton absorption. The energy exchange efficiency decreases gradually with the increasing of the relative time delay. With the positive time delay larger than 0.3 ps, the energy loss can be neglected because there is no strong coupling between the delay-tuned pulse and the two delay-fixed ones due to the decay of temporal overlap. Until the time delay of $\sim $ 40 ps, the energies of pulses almost recover to their initial levels. That is to say, the energy exchange in the domain of relative time delay has been elongated to the order of magnitude of 40 ps. This result was unexpected because no reported work has mentioned the process of energy exchange as a function of time delay beyond 3.0 ps. For instance, the duration of energy exchange in the relative time delay could be estimated as 400 fs in Ref. [9], and further Y. Liu \textit{et al} observed this time region of energy exchange was about 2.0 ps when two femtosecond laser pulses were involved \cite{PRL.105}. 

\begin{figure}[ht]
  \centerline{\includegraphics [clip, width=3.2in, angle=0]{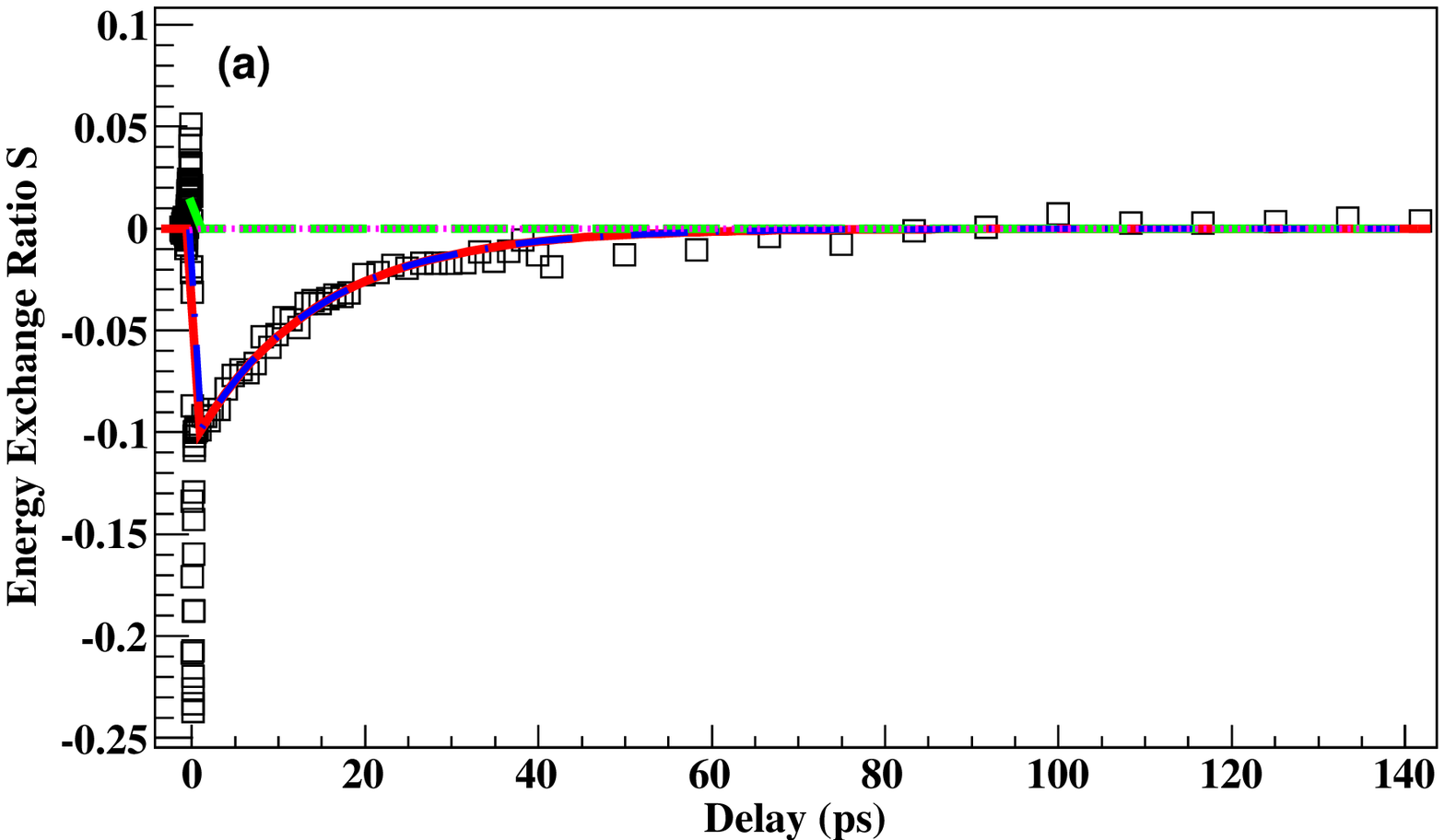}}
   \centerline{\includegraphics [clip, width=3.2in, angle=0]{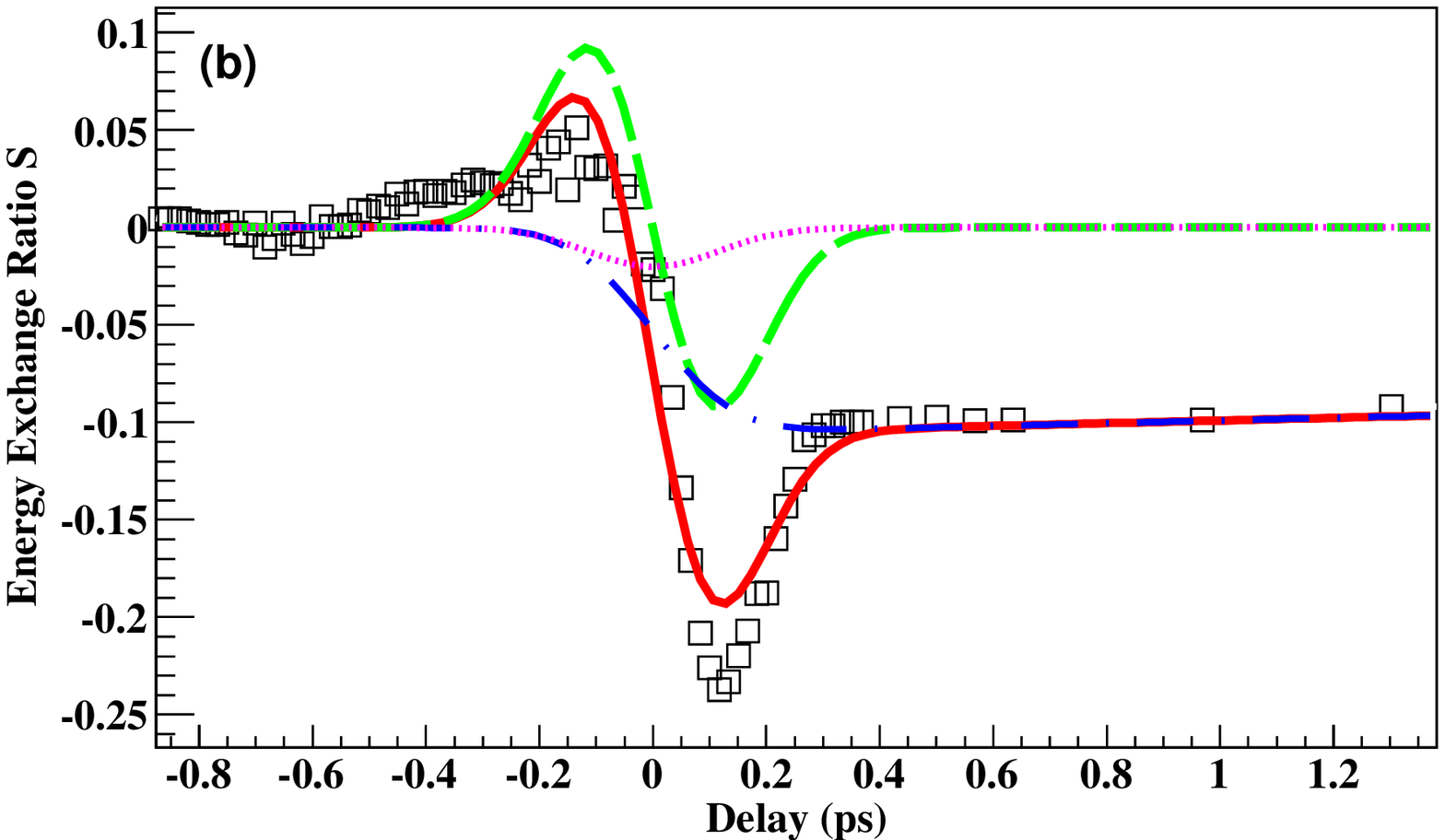}}  
  \caption{(Color online) (a) The energy exchange ratio S as a function of the relative time delay. The red solid (green dashed, blue dash-doted, violet dotted) line represents the best fit by using the formula $S(\tau )$ ($S_{1}(\tau )$, $S_{2}(\tau )$, $S_{3}(\tau )+S_{4}(\tau )$). (b) The detailed energy exchange in the relative time delay from $-0.8$ to 1.2 $ps$.}
  \label{FIG-3}
\end{figure}

Similar to previous work\cite{PRL.102}, the ratio of energy exchange between the delay-tuned pulse and the two delay-fixed ones can be defined as 
\begin{eqnarray}
 S = \frac{(E_{tuned}-E_{fixed}) - (E_{tuned_0}-E_{fixed_0})}{E_{tuned}+E_{fixed}}, 
\end{eqnarray}
where $E_{tuned}$ and $E_{fixed}$ represent the delay-tuned energy and energy of two delay-fixed pulses, $E_{tuned_0}$ and $E_{fixed_0}$ are the initial delay-tuned energy and total energy of two delay-fixed pulses, respectively. As shown in Fig. 3, the maximum energy exchange ratio can reach $5 \%$ when the time delay is negative, and $23 \%$ when the time delay is positive. When the delay-tuned energy gradually approach that of the two delay-fixed pulses, the two delay-fixed pulses transfer more energy to the delay-tuned one before the zero time delay, and gain less energy from the delay-tuned one after the zero time delay. Figure 4 shows the energy exchange ratio S as a function of the relative time delay when the delay-tuned beam orthogonally crossed the two delay-fixed beams. The direction of energy exchange reversed compared to that of small-angle crossing configuration, and there was no long-time dependence on the relative time delay. In this case, the interaction region was dramatically reduced, leading to a reduction of plasma density. Thus, the flow of energy transfer changed and the elongation of energy exchange in the delay domain vanished.

% vertical crossing scheme
\begin{figure}[ht]
  \centerline{\includegraphics [clip, width=3.2in, angle=0]{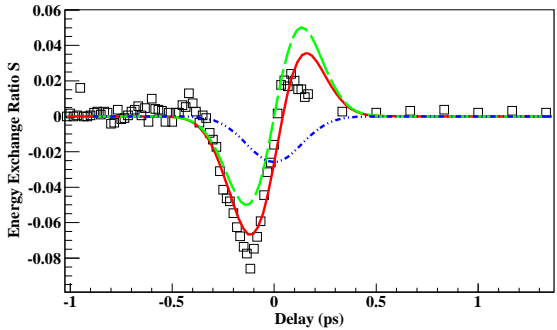}}
  \caption{(Color online) The energy exchange ratio S as a function of the relative time delay in the orthogonal crossing configuration. The red solid (green dashed, blue dash-doted) line represents the best fit by using the formula $S(\tau )$ ($S_{1}(\tau )$, $S_{3}(\tau )+S_{4}(\tau )$).}
  \label{FIG-4}
\end{figure}

% theoretical explanation and fit 
% Source: time-domain theory from N. Tang et al / J. Opt. Soc. Am. B 14, 3417
Because the electron density within plasma grating has been dramatically enhanced by using three filaments interaction, which can be confirmed by the significant enhancement of fluorescence intensity at the intersecting region, the plasma grating has a much larger decay time than the time scale of pulse overlap. X. Yang \textit{et al} has demonstrated that the plasma grating formed by two 45 fs filaments could last for about ~30 ps \cite{APL.97}. In addition, L.P. Shi \textit{et al} has demonstrated that the electron density in the plasma grating formed by two UV filaments decays exponentially, and its lifetime is on the order of 100 ps \cite{PRL.107}. The plasma grating gradually decays and consequently disappears due to two distinct processes \cite{PRE.86}, e.g. ambipolar diffusion and electron recombination. The slower the plasma decay process, the larger the imbanlance between the positive and negative portions of the energy exchange. Therefore, we can expect the long-time dependence of energy exchange on the relative time delay, corresponding to the lifetime of plasma grating formed by the noncollinear filaments interaction.

We use the treatment developed by N. Tang \textit{et al} \cite{J.OPT.SOC.AM.14.3412} to fit the energy exchange ratio, shown as solid line in Fig. 3(a) and 3(b). Taking the temporal response function $ R(t) $ of air as $exp(-t/\tau _{n})$, where $\tau _{n}$ is the exponential decay constant \cite{PRL.102}, the energy exchange ratio can be estimated as:
\begin{eqnarray}
\begin{aligned}
 S(\tau) = S_{1}(\tau ) + S_{2}(\tau ) + S_{3}(\tau ) + S_{4}(\tau ),
\end{aligned}
\end{eqnarray}

\begin{eqnarray}
\begin{aligned}
 S_{1}(\tau ) = \kappa \frac{\sqrt{\pi } \tau _{p} }{I_{0}^{2}} B_{xxxx} \{ -Im [  \int _{-\infty }^{+\infty} dt u^{*}(t-\tau ) u(t) \\
 \times \int _{-\infty }^{t} dt^{'} R(t-t^{'}) u(t^{'}-\tau ) u^{*}(t^{'}) ] \},
\end{aligned}
\end{eqnarray}

\begin{eqnarray}
\begin{aligned}
 S_{2}(\tau ) = \kappa \frac{\sqrt{\pi } \tau _{p} }{I_{0}^{2}} B_{xxxx}^{'} \{ -\int _{-\infty }^{+\infty} dt | u(t-\tau ) |^{2} \\
 \times \int _{-\infty }^{t} dt^{'} R(t-t^{'}) | u(t^{'}) |^{2} \},
\end{aligned}
\end{eqnarray}

\begin{eqnarray}
\begin{aligned}
 S_{3}(\tau ) = \kappa \frac{\sqrt{\pi } \tau _{p} }{I_{0}^{2}} B_{xxxx}^{'} \{ -Re [ \int _{-\infty }^{+\infty} dt u^{*}(t-\tau ) u^{*}(t) \\ 
 \times \int _{-\infty }^{t} dt^{'} R(t-t^{'}) u(t^{'}-\tau ) u(t^{'}) ] \},
\end{aligned}
\end{eqnarray}

\begin{eqnarray}
\begin{aligned}
 S_{4}(\tau ) = \kappa \frac{\sqrt{\pi } \tau _{p} }{I_{0}^{2}} B_{xxxx}^{'} \{ -Re [ \int _{-\infty }^{+\infty} dt u^{*}(t-\tau ) u(t) \\ 
 \times \int _{-\infty }^{t} dt^{'} R(t-t^{'}) u(t^{'}-\tau ) u^{*}(t^{'}) ] \},
\end{aligned}
\end{eqnarray}
where $\tau _{p}$ the pulse duration, $u(t)$ the time-dependent electric field, and $I_{0}=\int _{-\infty }^{+\infty} |u(t)|^{2}dt $. $S_{1}(\tau )$ represents the contribution from beam-coupling, and the terms of $S_{2}(\tau )$, $S_{3}(\tau )$ and $S_{4}(\tau )$ represent the nonlinear absorptive contributions. $\kappa=24 \sqrt{\pi } k L_{eff} E_{p} / n^{2} w^{2} c \tau _{p}$, where $k$ the wave number for the incident laser wavelength in vacuum, $L_{eff}$ the effective propagation length (3 cm), $E_{p}$ the pulse energy, and $c$ the light speed. The Gaussian beam radius $w$ at the intersection region was estimated as $100 \ \mu m$. $B_{xxxx}$ and $B_{xxxx}^{'}$ are the delayed contributions to the nonlinear refractive effects and the nonlinear absorptive effects respectively. Though these expressions are derived from the time-domain theory for pump-probe experiment, it is still valid for the situation involving three noncollinear filamentary beams, because the Beam-2 and Beam-3 were kept spatiotemporally overlapping.

The red solid lines in Fig. 3(a) and 3(b) were the best fit curves by using Eq. (2-6) with the relevant pulse duration and energies, and with a fit of $B_{xxxx} = 2.36 \times 10^{-7} \ cm^{3} / erg \cdot s$, $B_{xxxx}^{'} = 2.58 \times 10^{-7} \ cm^{3} / erg \cdot s$ and $\tau _{n} = 14.2 \ ps $. The third-order nonlinear susceptibility $\chi ^{(3)}$ can be estimated by $\chi ^{(3)} \approx 3 B_{xxxx} \tau _{n} = 1.006 \times 10^{-17} \ cm^{3}/erg $, and then the delayed nonlinear refractive index coefficient $n_{2,d} = 12 \pi ^{2} \chi ^{(3)} / n_{0}^{2} c = 3.97 \times 10^{-19} \ cm^{2}/W $, which is close to the reported value of $\sim 3.01 \times 10^{-19} \ cm^{2}/W $ in literature\cite{J.Opt.Soc.Am.B.14.650, APL.99}. The fit in Fig. 4 obtained $n_{2,d} = 2.20 \times 10^{-19} cm^{2}/W $, which was also in the same order of magnitude. Despite the complicated situation involving three filamentary pulses, the model mentioned above predict well the temporal evolution of energy exchange for the intersecting filamentary pulses. A detailed explanation will require full knowledge of the temporal and spatial evolution of the filamentary pulses at the pulse overlap region. 

% \section{Summary}
In summary, we have demonstrated the elongation of energy exchange process in the relative time delay between two delay-fixed pulses and a delay-tuned one. Corresponding to the gradual decay of plasma density, energy exchange exhibits a long-time dependence on the relative time delay. In the orthogonally crossing configuration, the flow direction of energy transfer reversed because of the reduction of plasma density. By using a time-domain model including the nonlinear absorptive effects, the best fit was obtained and the delayed nonlinear refractive index coefficient was extracted. These results indicate that efficient energy exchange between femtosecond pulses could be ensured and controlled by plasma formation assisted by filaments interaction. Further experiments will concentrate on control of filament onset and find new configuration of efficient energy exchange.

\vspace{3mm}
% acknowledgments
We would like to acknowledge the support of the National Natural Science Foundation of China (Grants No. 11135002,
No. 11075069, No. 91026021, No. 11075068, and No. 11175076), and a Scholarship Award for Excellent Doctoral Student granted by the Ministry of Education. 

% references

%\nocite{*}
% \bibliography{}  Produces the bibliography via BibTeX.

\end{document}